\documentclass[conference]{IEEEtran}
\IEEEoverridecommandlockouts
\usepackage{cite}
\usepackage{amsmath,amssymb,amsfonts}
\usepackage{algorithmic}
\usepackage{graphicx}
\usepackage{textcomp}
\usepackage{xcolor}
\usepackage{soul}
\usepackage{booktabs}
\usepackage[normalem]{ulem}
\usepackage{booktabs}
\usepackage{threeparttable}
\usepackage{stfloats}
\usepackage{multirow}
\usepackage{colortbl}
\usepackage{makecell}
\usepackage{amssymb}
\newcommand{\cmark}{\checkmark}
\usepackage{svg}
\usepackage{pifont}
\usepackage{amssymb} 
\usepackage{orcidlink}
\usepackage{cleveref}
\newcommand{\xmark}{\ding{55}} 
\usepackage[a4paper, total={184mm,239mm}]{geometry}
\def\BibTeX{{\rm B\kern-.05em{\sc i\kern-.025em b}\kern-.08em
    T\kern-.1667em\lower.6ex\hbox{E}\kern-.125emX}}
\usepackage[acronym]{glossaries}
\glsenableentrycount
\newacronym{lora}{LoRA}{Low-Rank Adaptation}
\newacronym{ai}{AI}{Artificial Intelligence}
\newacronym{cnn}{CNN}{Convolutional Neural Network}
\newacronym{cct}{CCT}{Compact Convolutional Transformer}
\newacronym{dma}{DMA}{Direct Memory Access}

\newacronym{fp}{FP}{floating point}
\newacronym{fpu}{FPU}{Floating Point Unit}
\newacronym{gemm}{GEMM}{General Matrix Multiplication}
\newacronym{gmacs}{GMACs}{Giga multiply–accumulate operations}

\newacronym{iot}{IoT}{Internet of Things}
\newacronym{lp}{LP}{Linear Probing}
\newacronym{l1}{L1}{Level-1 Tightly-Coupled Data Memory (TCDM)}
\newacronym{l2}{L2}{L2 On-Chip SRAM}
\newacronym{l3}{L3}{L3 External Memory (e.g., HyperRAM)}
\newacronym{tcdm}{TCDM}{Tightly-Coupled Data Memory}

\newacronym{mcu}{MCU}{Microcontroller Unit}
\newacronym{mlp}{MLP}{Multi-Layer Perceptron}
\newacronym{mops}{MOps}{Million operations}
\newacronym{peft}{PEFT}{Parameter-Efficient Fine-Tuning}
\newacronym{ffn}{FFN}{Feed-Forward Network}

\newacronym{riscv}{RISC-V}{Reduced Instruction Set Computer V}
\newacronym{onnx}{ONNX}{Open Neural Network Exchange}
\newacronym{gvsoc}{GVSoC}{event-driven virtual SoC simulator for PULP platforms}

\newacronym{relu}{ReLU}{Rectified Linear Unit}
\newacronym{redmule}{RedMulE}{Reduced Multiplier Engine (FP GEMM accelerator)}
\newacronym{sgd}{SGD}{Stochastic Gradient Descent}
\newacronym{soc}{SoC}{System-on-Chip}
\newacronym{tinyml}{tinyML}{tiny Machine Learning}

\usepackage[font=small]{caption}
\begin{document}
\title{TrainDeeploy: Hardware-Accelerated Parameter-Efficient  Fine-Tuning of Small Transformer Models at the Extreme Edge 
}

\author{
\IEEEauthorblockN{
    Run Wang\IEEEauthorrefmark{1}~\orcidlink{0000-0002-1011-6432},
    Victor J.B. Jung\IEEEauthorrefmark{1}~\orcidlink{0009-0001-7462-3468},
    Philip Wiese\IEEEauthorrefmark{1}~\orcidlink{0009-0001-7214-2150},
    Francesco Conti\IEEEauthorrefmark{2}~\orcidlink{0000-0002-7924-933X},
    Alessio Burrello\IEEEauthorrefmark{3}~\orcidlink{0000-0002-6215-8220},
    Luca Benini\IEEEauthorrefmark{1}\IEEEauthorrefmark{2}~\orcidlink{0000-0001-8068-3806}
    \thanks{Accepted at the Design, Automation and Test in Europe (DATE) 2026 conference. 
This is the author's version of the work. The final version will appear in IEEE Xplore.}
}
\IEEEauthorblockA{
    \IEEEauthorrefmark{1}\textit{Integrated Systems Laboratory (IIS), ETH Zurich}, Switzerland \\
    \IEEEauthorrefmark{2}\textit{Department of Electrical, Electronic and Information Engineering (DEI), University of Bologna}, Italy \\
    \IEEEauthorrefmark{3}\textit{Department of Control and Computer Engineering (DAUIN), Politecnico di Torino}, Italy
}
\IEEEauthorblockA{
    \{runwang, jungvi, wiesep, lbenini\}@iis.ee.ethz.ch, f.conti@unibo.it, alessio.burrello@polito.it
  }
    \vspace{-0.5cm}
}
  
\maketitle
\begin{abstract}
On-device tuning of deep neural networks enables long-term adaptation at the edge, while keeping data fully private and secure. However, the high computational demand of backpropagation remains a challenge for ultra-low-power, memory-constrained extreme-edge devices. Attention-based models further exacerbate this challenge, given their complex architecture and scale. We present \textit{TrainDeeploy}, a novel framework that unifies efficient inference with on-device training on heterogeneous ultra-low-power \glspl{soc}. TrainDeeploy is the first complete on-device training pipeline for extreme edge \glspl{soc} supporting both \glspl{cnn} and Transformer models, as well as multiple training techniques, such as selective layer-wise fine-tuning and \gls{lora}. On a RISC-V-based heterogeneous \gls{soc}, we demonstrate the first end-to-end fine-tuning of a complete Transformer,  \gls{cct}, achieving 11 trained images per second. We show that \gls{lora} on-device leads to a 23\% reduction in dynamic memory usage, $15\times$ reduction in trainable parameters and gradients, and $1.6\times$ reduction in memory transfer compared to full backpropagation. TrainDeeploy achieves up to 4.6\,FLOP/cycle on CCT (0.28M Param, 71–126M FLOPs) and leading-edge performance up to 13.4\,FLOP/cycle on Deep-AE (0.27M Param, 0.8M FLOPs), while simultaneously widening the scope compared to state-of-the-art frameworks to support both CNNs and Transformers with parameter-efficient tuning.
\end{abstract}



\glsresetall

\begin{IEEEkeywords}
On-device Training, Edge AI, Hardware Acceleration, LoRA, Heterogeneous Platforms
\end{IEEEkeywords}

\section{Introduction}
Edge computing has emerged as a key enabler of intelligent \cgls{iot} systems, enabling data processing and decision-making directly on end devices~\cite{lin_tiny_2023}. With the rapid integration of \cgls{ai}, this paradigm has extended from mobile devices to \cgls{mcu}-based \cglspl{soc} powering sensors and wearables, bringing deep learning capabilities to ultra-low-power hardware~\cite{reuther_ai_2022}. Considerable progress has been made in deploying \cglspl{cnn} and Transformers on such platforms, demonstrating that even highly resource-constrained devices can support efficient inference~\cite{scherer_deeploy_2024}. However, the potential of \cgls{ai} at the edge extends beyond inference. Growing demand for personalization, privacy, and continual adaptation motivates on-device training~\cite{zhu_-device_2024}, which typically relies on backpropagation to fine-tune pretrained models locally without reliance on the cloud. Alternative approaches, such as gradient-free~\cite{hinton_forward-forward_2022, zhang_revisiting_2024} or meta-learning~\cite{blanken_chameleon_2025} methods, have been explored, but they lack the generality of backpropagation and, in many cases, cannot achieve competitive accuracy results.

On the other hand, training \cglspl{cnn} and Transformer models with backpropagation is highly demanding in terms of both computation and memory, particularly when facing the hardware constraints of extreme edge devices. 
Even compact networks typically require $10^7$–$10^9$ floating-point operations per training step, as the workload is dominated by large \gls{gemm} operations~\cite{lin_-device_nodate}.
Moreover, storing activations for gradient computation requires more than $10^7$ bytes of memory~\cite{lin_tiny_2023, lin_-device_nodate}, which often exceeds the capacity of embedded platforms.

Current embedded machine learning training frameworks address these challenges via compute-focused optimization, parameter-efficient methods, or lightweight \cgls{cnn}-centric designs, but each has clear limitations. For example,  PULP-TrainLib~\cite{orailoglu_pulp-trainlib_2022} primarily targets compute performance optimization; however, it does not provide an end-to-end flow with operator tiling and memory allocation across multiple hierarchies, leading to inefficient memory usage. Pruning and sparse training techniques such as MiniLearn~\cite{profentzas_minilearn_nodate}, TTE~\cite{lin_-device_nodate}, Sparse Backpropogation~\cite{paissan2024structured} reduce the memory demand, but they remain largely CNN-centric and are tailored to single\nobreakdash-core MCUs.

To address these challenges, we introduce \textit{TrainDeeploy}, a compilation and execution
flow that enables Transformer training on heterogeneous ultra\nobreakdash-low\nobreakdash-power \glspl{soc}. Building on Deeploy~\cite{scherer_deeploy_2024}, which was created for energy-efficient inference on heterogeneous \glspl{mcu}, TrainDeeploy extends the flow with automatic differentiation and training-oriented passes, making backpropagation feasible on heterogeneous resource-constrained platforms.
We demonstrate effective on-device Transformer fine-tuning on a heterogeneous \cgls{soc} with a \cgls{gemm} accelerator, using \cgls{lora}~\cite{hu_lora_2021} to reduce the memory footprint.
This establishes the first unified pipeline for both inference and training of Transformers on ultra-low-power heterogeneous \cglspl{soc}.

Specifically, our contributions are summarized as follows:
\begin{itemize}
    \item We present TrainDeeploy, a compilation and execution flow that enables Transformer training on heterogeneous ultra\nobreakdash-low\nobreakdash-power \glspl{soc}, leveraging on-chip accelerators to speed up training.
    \item We demonstrate the first complete end-to-end on-device fine-tuning of a \gls{cct} on a heterogeneous \cgls{soc} platform~\cite{prasad_siracusa_2024}.
    \item We implement on-device \gls{lora} training as a technique to reduce on-device training workload, making it feasible on low-power extreme-edge devices. 

\end{itemize}

Our results, targeting an extreme-edge RISC-V-based heterogeneous SoC with on-chip GEMM acceleration, show that \cgls{lora} acceleration yields $23\%$ lower dynamic memory usage, $15\times$ fewer trainable parameters and gradients, $1.6\times$ fewer memory transfers, and speeds up the training by 2.3-3.5x compared to non-accelerated execution, while also outperforming state-of-the-art \cgls{cnn} training frameworks. We show that end-to-end \gls{cct} fine-tuning can be performed at a throughput of up to 11 gradient updates per second in a single-sample training setting, fine-tuning all transformer layers. To the best of our knowledge, TrainDeeploy is the first end-to-end on-device fine-tuning deployment framework targeting both CNNs and Transformers.
\vspace{-0.6em}
\section{Background}
\label{sec:background}

\subsection{On-Device Training}
On-device training imposes substantially higher demands compared with inference in terms of computational complexity, memory footprint, and numerical precision. Beyond forward propagation, training requires backpropagation to compute gradients for both inputs and weights. The most important operation is the \cgls{gemm}, given its massive utilization in attention layers, \cglspl{mlp}, and \cglspl{cnn}. For a generic \cgls{gemm} $Y = W X$, the backward pass yields
$\tfrac{\partial L}{\partial X} = W^{\top} \Delta$ and 
$\tfrac{\partial L}{\partial W} = \Delta X^{\top}$, 
where $\Delta = \tfrac{\partial L}{\partial Y}$ is the upstream gradient.
 Each forward GEMM, therefore, induces two additional GEMMs in the backward pass. In addition to computation, intermediate activations must be stored until the backward pass, significantly stressing the few hundred KB to a few MB of SRAM available in typical \cgls{mcu}-class devices. Furthermore, while inference commonly tolerates low-precision integer quantization~\cite{lin_tiny_2023}, training generally requires higher-precision \cgls{fp} arithmetic (e.g., FP16, FP32) to ensure stable gradient updates~\cite{lin_-device_nodate}. 

\subsection{Parameter-Efficient Fine-Tuning and LoRA}
\begin{figure}
    \centering
    \includegraphics[width=\linewidth]{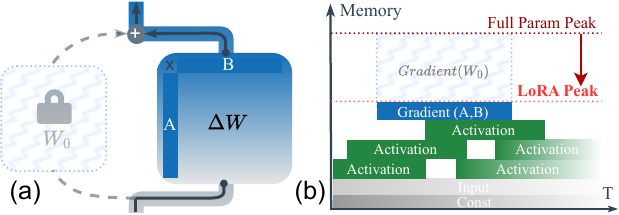}
    \caption{Low-Rank Adaptation (\cgls{lora}). 
    (a) Frozen pre-trained weight $W_0$ with trainable low-rank matrices $A$ and $B$.
    (b) Comparison of memory footprint over time between full-parameter fine-tuning and \cgls{lora}. 
The stacked areas illustrate how tensors are allocated and released during execution. By reducing gradient storage, \cgls{lora} further lowers the peak memory footprint, enabling training within tight on-chip memory budgets.}
    \label{fig:lora}
    \vspace{-0.3cm}
\end{figure}

\cgls{peft} techniques, which update only a small subset of model parameters while retaining accuracy and adaptability, have been introduced to cope with the prohibitive compute and memory requirements of billion-parameter large language models. Despite the orders-of-magnitude difference in scale, such techniques are also highly relevant but remain largely unexplored in the context of MCU-scale, edge-oriented platforms. Among these methods, \cgls{lora} is particularly effective~\cite{hu_lora_2021}. Instead of updating the full weight $W_0 \in \mathbb{R}^{d \times k}$, \cgls{lora} introduces a low-rank decomposition $W = W_0 + BA$, where $B \in \mathbb{R}^{d \times r}$, $A \in \mathbb{R}^{r \times k}$, and $r \ll \min(d,k)$. As shown in \Cref{fig:lora}(a), $W_0$ is frozen and only the two small matrices $A$ and $B$ are trained, reducing the number of trainable parameters from $dk$ to $r(d+k)$. \Cref{fig:lora}(b) illustrates how this translates into a substantial reduction in memory footprint compared to full-parameter fine-tuning, with activation and weight memory remaining unchanged while gradient storage is drastically reduced. Also, since gradients are computed only for $A$ and $B$, \cgls{lora} leaves the activation-gradient computation unchanged but significantly reduces the weight-gradient and parameter-update costs.
\vspace{-0.2em}
\subsection{Heterogeneous HW Platforms}
\label{sec:HW_heterogeneous}
A recent trend in the extreme-edge AI domain is the proliferation of heterogeneous \cglspl{soc} equipped with specialized accelerators. Recent platforms increasingly combine a general-purpose host with neural processing units (NPUs), digital signal processors (DSPs), or fixed-function GEMM engines. For example, the STM32N6 family from STMicroelectronics integrates the Arm Cortex-M55 core with the in-house Neural-ART accelerator~\cite{stm32n6}. The MAX78000 from Analog Devices features an Arm Cortex-M4 control core coupled with a dedicated \cgls{cnn} accelerator~\cite{max78000}. The GAP9 from GreenWaves Technologies consists of a RISC-V fabric controller and a compute cluster with nine RISC-V cores plus the NE16 deep neural network accelerator~\cite{gap9}. Finally, Arm Ethos-U55 NPUs are integrated in commercial edge \cglspl{soc} such as the Alif Ensemble family, Infineon PSoC Edge, and Himax WiseEye2~\cite{ethos_u55}. This heterogeneous organization enables collaborative execution: control and irregular tasks remain on the MCU core, while specific compute-intensive tasks are offloaded to accelerators. 

\section{Related Work}
\label{sec:related}
Prior work on on-device training for extreme-edge platforms has explored four complementary directions. Model-side methods (e.g., TinyTL~\cite{cai_tinytl_2021}, MiniLearn~\cite{profentzas_minilearn_nodate}) reduce memory by pruning or restricting updates, but sacrifice adaptability and remain CNN-centric. Compiler-level techniques (e.g., TTE~\cite{lin_-device_nodate}, POET~\cite{patil_poet_2022}) mitigate memory pressure via operator reordering, sparsity, quantization, paging, or gradient checkpointing, but require nontrivial graph modifications and some tricks incur latency–memory trade-offs.
System-level frameworks such as PULP-TrainLib~\cite{orailoglu_pulp-trainlib_2022} offer optimized computational primitives for on-device RISC-V learning, but mainly demonstrate benefits on small networks and lack memory-aware compilation support. Hardware design for extreme edge training (e.g., MINOTAUR~\cite{prabhu_minotaur_2025}, Chameleon~\cite{blanken_chameleon_2025}) achieves energy-efficiency via novel formats or gradient-free schemes, but limits generality and portability. Overall, these approaches often \textit{(i)} compromise accuracy, \textit{(ii)} optimize latency or memory in isolation, \textit{(iii)} remain CNN-centric, or \textit{(iv)} depend on specialized formats and hardware. To the best of our knowledge, no work has demonstrated an end-to-end Transformer/CNN training pipeline within the energy and memory budgets of ultra-low-power devices, one of the core contributions of this work.
Furthermore, none of these tools support the efficient deployment on highly heterogeneous devices, as the ones described in Sec. \ref{sec:HW_heterogeneous}. TrainDeeploy is the first end-to-end Transformer/CNN training pipeline for heterogeneous ultra-low-power devices.

\begin{figure*}[t] 
    \centering
    \includegraphics[width=\linewidth]{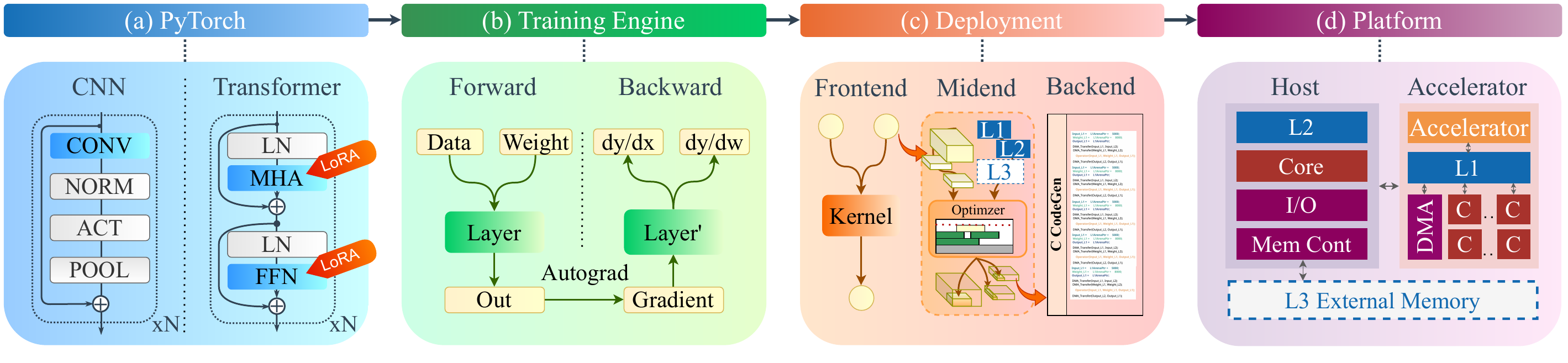}
    \caption{ Overview of the TrainDeeploy framework. 
    (a) Models are defined in PyTorch and exported through \cgls{onnx}. 
    (b) The training engine augments the forward graph with the backward graph via autograd, producing a full training graph. 
    (c) Deeploy extends its inference compiler to training with a frontend (\cgls{onnx} parsing), a midend (memory optimizer integrating tiling and static allocation across full forward-backward graph), and a backend (C code generation). 
    (d) The generated code is deployed on the heterogeneous \cgls{soc}leveraging hierarchical memory mapping (\cgls{l1}--\cgls{l3}) and an on-board accelerator for \cglspl{gemm}.
    }
    \label{fig:methodology_overview}
    \vspace{-0.3cm}
\end{figure*}

\section{TrainDeeploy Framework}
\label{sec:framework}

Starting from PyTorch models and training strategies, TrainDeeploy performs training graph construction, operator tiling, and static memory allocation at compile time and lowers the graph into C code, which can then be compiled using a regular platform compiler (e.g., GCC, LLVM). To realize this, we build on Deeploy \cite{scherer_deeploy_2024}, a domain-specific compiler aimed at energy-efficient inference of DNNs on heterogeneous MCUs. TrainDeeploy exploits and extends Deeploy’s static memory allocation and tiling principles, broadening the flow from inference to training. We also designed our framework to take advantage of accelerators that are increasingly available in emerging edge devices. Within this context, a central contribution is the joint handling of memory and compute bottlenecks: \cgls{lora} reduces trainable state and gradient pressure to fit the HW memory, while we offload GEMM workloads to on-chip accelerators to speed up training workloads.

\paragraph{Target platform}
As \Cref{fig:methodology_overview}(d) illustrates, the target platform we consider is a standard heterogeneous SoC organization, consisting of a host processor that manages peripherals and orchestrates execution, and heterogeneous programmable accelerators. Specifically, we adopt a generic accelerator paradigm that might contain one or more processor cores, a local L1 memory, and potentially a fixed-function hardware unit (e.g., to accelerate GEMM). The memory hierarchy consists of on-chip memories, i.e., an accelerator-dedicated L1, a host L2, and an external L3, with data movement managed with DMA.  

\paragraph{Pipeline} 
The pipeline starts with the input models and training strategies, which are defined in PyTorch (\Cref{fig:methodology_overview}(a)). While our framework supports general architectures such as CNNs and Transformers, we highlight that Transformer networks can optionally be augmented with \cgls{peft} modules such as \cgls{lora}. More broadly, TrainDeeploy can integrate other model-level training optimizations (e.g., structured sparsity, layer-wise training, or alternative PEFT methods). The training configuration, loss functions, and optimizers are specified, and the models are then exported in \cgls{onnx} format to provide a hardware-agnostic graph representation that serves for subsequent symbolic differentiation.  

\Cref{fig:methodology_overview}(b) illustrates the construction of a full training graph from its forward counterpart. Using an automatic differentiation engine~\cite{onnxruntime_github}, the forward graph is traversed in reverse topological order starting from the loss node. For each operator, a predefined differentiation rule specifies the corresponding backward operator, which is instantiated and linked to form explicit gradient paths. The resulting structure integrates the forward and backward dataflows into an operator-level automatic differentiation graph. Subsequently, optimizer update rules are incorporated as dedicated subgraphs, yielding a complete training representation. The final outcome is a static ONNX training graph that jointly encodes inference and gradient flows in a hardware\nobreakdash-agnostic intermediate form. Unlike dynamic autograd execution, the static representation exposes the complete forward\nobreakdash–backward structure of the training step. This global view enables the subsequent memory optimizer to operate over the entire computation, thereby enlarging the search space.

\Cref{fig:methodology_overview}(c) shows the graph compiler, the core of TrainDeeploy, which processes the training graph with awareness of the memory hierarchy of heterogeneous SoCs. The frontend firstly parses and validates operators to corresponding kernels before lowering them into a static intermediate representation. In the Midend, memory scheduling is performed in conjunction with operator tiling through a unified constraint\nobreakdash-programming formulation. Tiling constraints, derived from kernel\nobreakdash-specific requirements and hardware limits, define the feasible search space for tile sizes. At the same time, the scheduler applies liveness analysis to model tensor lifetimes. These two aspects are solved jointly as a 2D bin\nobreakdash-packing problem, producing a static allocation schedule using TetriSched~\cite{tumanov2016tetrisched} that minimizes peak memory usage across the hierarchy while ensuring feasibility for all operators of the full forward\nobreakdash-backward graph. By default, L1 stores active tiles, while L2 holds weights, inputs, activations, and gradients; when the L2 budget is exceeded, tensors spill to L3. \cgls{lora} low-rank matrices are treated as standard \cglspl{gemm} integrated seamlessly into the pipeline. Finally, the backend generates C code with optimized FP kernels for execution on the target platform, as illustrated in \Cref{fig:methodology_overview}(d), with accelerator-specific extensions discussed in the following paragraph.

\paragraph{Accelerator Support}
TrainDeeploy extends Deeploy with end-to-end compiler support for on-board \cgls{fp} \cgls{gemm} accelerators. The frontend identifies \cglspl{gemm} and convolutional layers that can be supported by the accelerator, the midend applies accelerator-aware tiling strategies to respect \cgls{l1} constraints, and the backend generates runtime kernels that manage synchronization with CPU operators and perform data layout transformations. By mapping \cgls{gemm}-heavy kernels from both native \cglspl{gemm} and lowered convolutions in forward and backward passes to the accelerator, TrainDeeploy effectively exploits on-chip compute units and alleviates the bottleneck of \cgls{gemm}-dominated training computations.

\section{TrainDeeploy Testing Scenario}
\subsection{Network Overview and Fine-Tuning Strategies}

\begin{figure}[tb]
    \centering
    \includegraphics[width=1\linewidth]{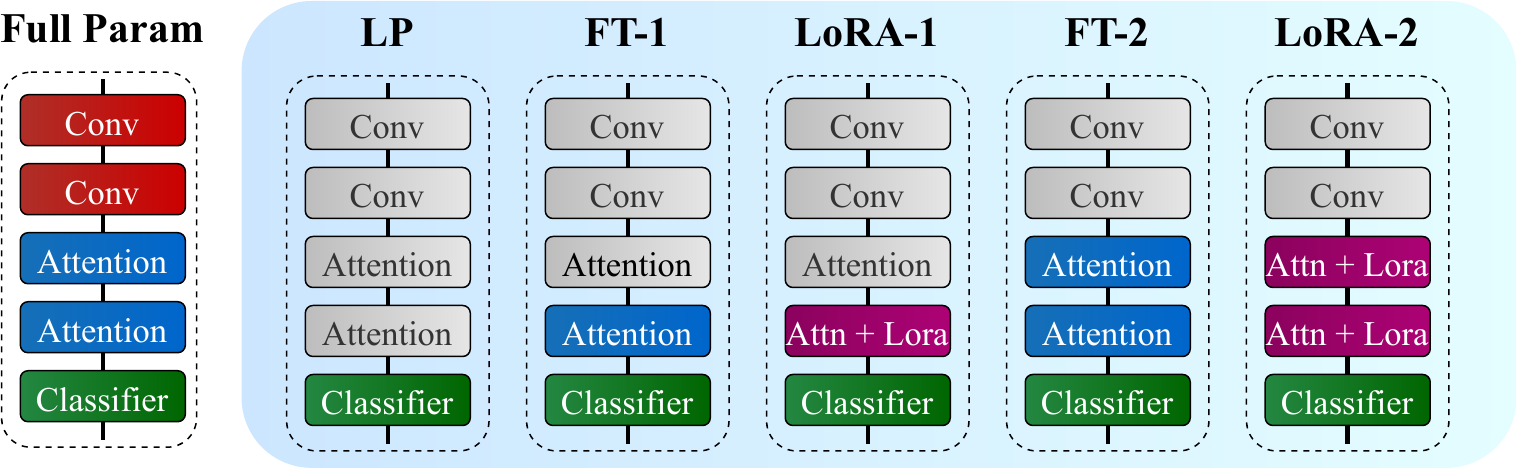}
    \caption{Illustration of fine-tuning strategies evaluated on the CCT-2 model for on-device training. 
    The convolutional tokenizer (\textbf{Conv}) is frozen in all strategies, while different subsets of transformer encoder blocks (\textbf{Attn}) and the classifier head are adapted. 
    Five representative strategies are considered: 
    \textbf{LP} (linear probing), only the classifier head is trained; 
    \textbf{FT-1}, full fine-tuning of the last attention block; 
    \textbf{LoRA-1}, low-rank adaptation (LoRA, rank $r=4$) applied to the last attention block; 
    \textbf{FT-2}, full fine-tuning of the last two attention blocks; 
    and \textbf{LoRA-2}, LoRA (rank $r=4$) applied to the last two attention blocks.}
    \label{fig:setup}
    \vspace{-0.3cm}
\end{figure}

We adopt the Compact Convolutional Transformer (CCT-2/3x2)~\cite{hassani_escaping_2022} 
as the target model to show the performance of TrainDeeploy. This lightweight vision transformer contains two convolutional tokenizer 
layers and two transformer encoder blocks (2 heads, 128-dimensional embedding, 
128-dimensional hidden \cgls{mlp}), followed by attention-based sequence pooling. 
Overall, it has 0.28\,M parameters ($\approx$1.12\,MB in FP32), requires only $67\,\mathrm{MFLOPs}$ per inference, 
and achieves 89.75\% accuracy on CIFAR-10 and 66.93\% on CIFAR-100.  

To evaluate the configurability of our framework, we freeze the convolutional tokenizer and consider five representative fine-tuning strategies, as shown in \Cref{fig:setup}. 
The first is linear probing (LP), where all transformer blocks remain frozen and only the classifier head is trained, serving as the most lightweight baseline. 
The second one, FT\nobreakdash-1, unfreezes the last attention 
block for full fine-tuning, while the third one, LoRA\nobreakdash-1,  applies rank-4 low-rank adaptation to the last attention block.
Similarly, FT\nobreakdash-2 unfreezes the last two attention blocks, and LoRA\nobreakdash-2 applies rank-4 low-rank adaptation to these two blocks.  
\Cref{tab:all_strategies} summarizes the trainable components, LoRA usage, accuracy, and cost of each strategy.

All training experiments are deployed on the hardware platform using \cgls{sgd} with a batch size of 1 and single-step parameter updates, with all computations (weights, activations, gradients, and optimizer states) in FP32.


\vspace{-0.6em}
\subsection{On-Device Fine-Tuning Hardware Setup}

\begin{figure}[t]
    \centering
    \includegraphics[width=1\linewidth]{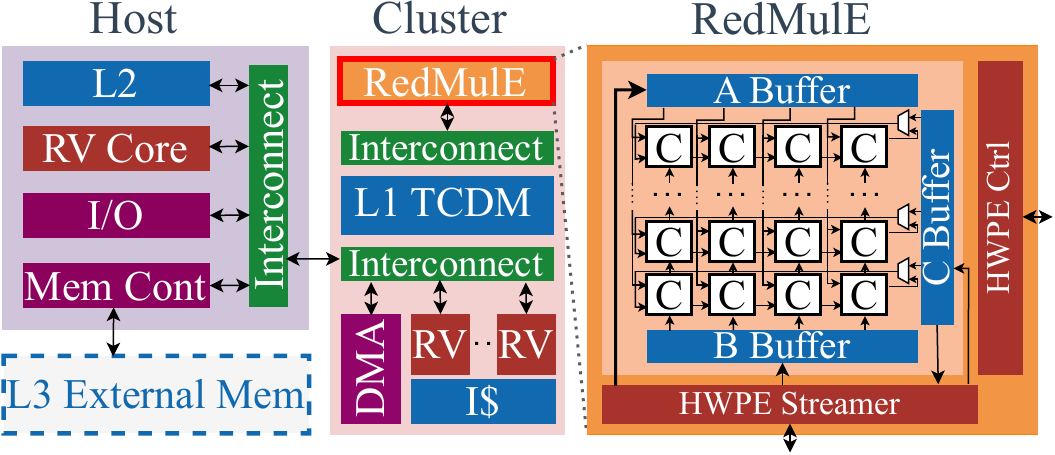}
  \caption{Hardware setup of the PULP-based SoC modeled in GVSoC. 
The system consists of a host and an 8-core compute cluster with shared multi-banked L1 TCDM (128\,KB), 
a hierarchical memory with L2 SRAM (2\,MB) and external L3 HyperRAM (32\,MB), 
and the \cgls{redmule} floating-point GEMM accelerator integrated with direct low-latency access to L1.}
\label{fig:hardware}
\vspace{-0.6cm}
\end{figure}
As a practical representative of the class of heterogeneous SoCs discussed in ~\Cref{sec:framework}, we selected a PULP instance simulated using the event-based GVSoC simulator~\cite{gvsoc_github}. As shown in \Cref{fig:hardware},
a host leverages a RISC-V core, while the accelerator cluster includes 8 RISC-V processors and a specialized GEMM unit based on the RedMulE architecture~\cite{tortorella_redmule_2023}. The RISC-V cores implement the \texttt{RV32IMFCXpulp} ISA and share four \cglspl{fpu}, 
all connected to a 128\,KB multi-banked L1, called \cgls{tcdm}, via a logarithmic interconnect for single-cycle, low-latency access. 
The memory hierarchy comprises 2\,MB L2 SRAM and 32\,MB external L3 HyperRAM accessible through a HyperBus interface. 
Since the combined peak memory usage of \cgls{cct} model weights, activations, and gradients exceeds the L2 capacity, 
L3 serves as the primary storage, with tensors staged through L2 and tiled into L1 at runtime.  
\cgls{redmule} is a \cgls{fp} GEMM engine based on a $12 \times 4$ systolic array of \cglspl{fpu} with a three-stage pipeline within a sub-100\,mW power envelope. It is originally optimized for FP16/FP8 inference workloads. In this work, the datapath is extended to FP32 to ensure stable gradient accumulation during training. \cgls{redmule} is tightly coupled to the RISC-V cluster through the \cgls{l1} \cgls{tcdm}. Based on the silicon prototype reported in~\cite{prasad_siracusa_2024}, we assume a target frequency of $360\,\mathrm{MHz}$ for the whole \cgls{soc}.

\vspace{-0.3em}
\section{Results}
\label{sec:results}
\subsection{Transfer Learning Results}

\subsubsection{Fine-Tuning Setup}
To show that our setup is credible for real-world applications, we report fine-tuning accuracy on two representative few-shot tasks: CIFAR-10 → MNIST and CIFAR-10 → EuroSAT, each under a 50-shot setting following the benchmark in~\cite{guo_broader_2020}.
Each model is trained for 100 epochs using \cgls{sgd} with a batch size of 8, 
an initial learning rate of 0.01, and cosine annealing down to 0.0005. 
The convolutional tokenizer is frozen, as it mainly provides generic low-level features, and unfreezing adds cost with little benefit. 
Final accuracy is reported as the average over the last five epochs, 
and each few-shot experiment is repeated 30 times for statistical robustness.  

\begin{table*}[t]
\centering
\caption{Comparison of fine-tuning strategies on CCT-2 and cost metrics. 
Accuracy is reported for 50-shot transfer (mean $\pm$ std over 30 runs). 
For reference, training from scratch yields 99.7\% on MNIST and 94.0\% on EuroSAT.}
\label{tab:all_strategies}
\scriptsize
\setlength{\tabcolsep}{4pt}
\renewcommand{\arraystretch}{0.9}
\resizebox{\textwidth}{!}{
\begin{tabular}{c c c|cc|cc}
\toprule
\multirow{2}{*}[-0.7ex]{\textbf{ \,\,\,\,Strategy \,\,\,\,}} &
\multirow{2}{*}[-0.7ex]{\textbf{Training Components \,\,\,\,}} &
\multirow{2}{*}[-0.7ex]{\textbf{LoRA \,\,\,\,}} &
\multicolumn{2}{c|}{\textbf{Accuracy (50-Shot Transfer)}} & 
\multicolumn{2}{c}{\textbf{Cost}} \\
\cmidrule(lr){4-5} \cmidrule(lr){6-7}
 &  &  & \textbf{ \,\,\,\,\,CIFAR-10$\rightarrow$MNIST \,\,\,} & \textbf{ \,\,\,CIFAR-10$\rightarrow$EuroSAT \,\,\,\,\,} & \,\,\,\,\,\textbf{ \,\,\,\,\,FLOPs (M)} & \textbf{ \,\,\,\,\, Trained Param(MB) \,\,\,\,\,} \\
\midrule
Full FT & Entire model             & \xmark & 92.83 $\pm$ 0.91 & 64.85 $\pm$ 3.11 & 201 & 1.12   \\
\midrule

LP      & Classifier head          & \xmark & 88.34 $\pm$ 0.42 & 76.70 $\pm$ 0.55 & 71  & 0.005 \\
FT-1    & Last attention block     & \xmark & 93.54 $\pm$ 0.38 & 78.94 $\pm$ 0.44 & 96  & 0.38 \\
\rowcolor{gray!15}
LoRA-1  & Last attention block     & \cmark & 95.38 $\pm$ 0.29 & 77.00 $\pm$ 0.47 & 86  & 0.026 \\
FT-2    & Last 2 attention blocks  & \xmark & 94.62 $\pm$ 0.33 & \textbf{81.52 $\pm$ 0.36} & 126 & 0.76 \\
\rowcolor{gray!15}
LoRA-2  & Last 2 attention blocks  & \cmark & \textbf{96.00 $\pm$ 0.27} & 80.50 $\pm$ 0.41 & 104 & 0.05 \\
\bottomrule
\vspace{-0.3cm}
\end{tabular}}
\end{table*}
\begin{figure}[t]
    \centering
    \includegraphics[width=1\linewidth]{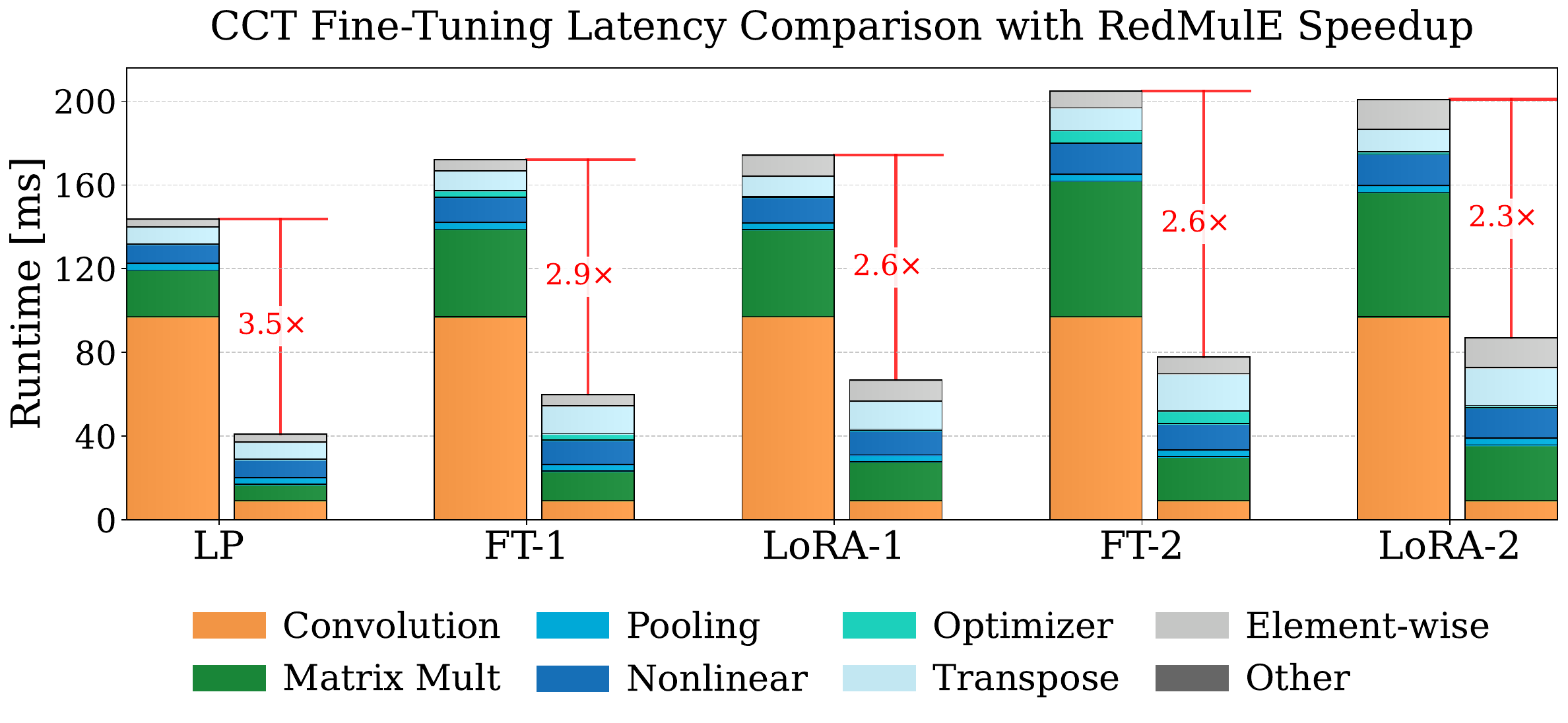}
    \caption{End-to-end training latency across fine-tuning strategies. 
    For each strategy, the left bar shows runtime using 8 cores without \cgls{redmule} acceleration, 
    while the right bar shows runtime with \cgls{redmule} acceleration. In the accelerated LoRA-2 and FT-2 configurations, the measured latency corresponds to a peak throughput of up to 11 gradient updates per second under single-sample, end-to-end fine-tuning.
    }
    \label{fig:hw_speedup}
    \vspace{-0.75cm}
\end{figure}

\subsubsection{Fine-Tuning Accuracy}
\Cref{fig:setup} and \Cref{tab:all_strategies} highlight the trade-off between accuracy and trainable footprint. 
As expected, linear probing yields the lowest accuracy, while adapting a single Transformer block (FT\nobreakdash-1) already improves transferability. 
Applying LoRA to the same block (LoRA\nobreakdash-1) further boosts MNIST accuracy to 95.4\%, showing that low-rank adaptation can enhance generalization while reducing parameters.  
Extending to two blocks (FT\nobreakdash-2) provides the highest EuroSAT accuracy (81.5\%), whereas LoRA\nobreakdash-2 achieves the best MNIST accuracy (96.0\%) with only 0.05\,MB trainable parameters, about $15\times$ fewer than FT\nobreakdash-2 for just 1\% accuracy gap on EuroSAT. 
For reference, training from scratch yields 99.7\% on MNIST and 94.0\% on EuroSAT.
Under the 50-shot setting, full fine-tuning of the entire model achieves 92.8\% ± 0.9 on MNIST and 64.8\% ± 3.1 on EuroSAT, exhibiting a substantial accuracy drop compared to configurations with a frozen convolutional tokenizer. Therefore, in the following experiments, we freeze the convolutional tokenizer and focus fine-tuning on the transformer layers.

\subsubsection{Training Cost}
As shown in \Cref{tab:all_strategies}, LP is computationally the cheapest, but it is insufficient in accuracy. 
Full layer fine-tuning incurs a steep cost (0.38–0.76\,MB), while LoRA keeps parameters small (0.026–0.05\,MB) and operations lower than full fine-tuning.  
This operation decrease stems from reducing gradient and optimizer states while leaving activations unchanged.  Overall, LoRA achieves nearly the same accuracy as full fine-tuning at a fraction of the cost, making Transformer adaptation increasingly practical within the tight memory and energy budgets of ultra-low-power \cglspl{soc}, even when considering future extensions to more complex optimizers and larger batch sizes.

\vspace{-0.3em}
\subsection{End-to-End Latency and Memory Footprint} 
\Cref{fig:hw_speedup} shows the end-to-end training runtime per sample per step at 360 MHz for the five fine-tuning strategies listed in \Cref{tab:all_strategies}. Each bar compares execution on the 8-core cluster (left) with \cgls{redmule} acceleration (right).
On the baseline system, runtimes range from 143 ms to 200 ms. LP is the fastest strategy since it updates only the last layers, while runtime increases as more layers are tuned. For the same number of updated layers, FT-1 and FT-2 are slower than their LoRA counterparts because they update more parameters. 

With \cgls{redmule} acceleration, runtimes drop to 41–87 ms, corresponding to a 2.3–3.5$\times$ speedup. LP achieves the highest gain (3.5$\times$) as it has the highest ratio of GEMM and convolution operations, while LoRA-2 shows the smallest improvement (2.3$\times$). Although runtime grows with the number of updated layers, LoRA can be slightly slower than FT after acceleration, due to its many small matrix multiplications limiting accelerator utilization and frequent low-rank transfers adding overhead.

\begin{figure}[t]
    \centering
    \includegraphics[width=1\linewidth]{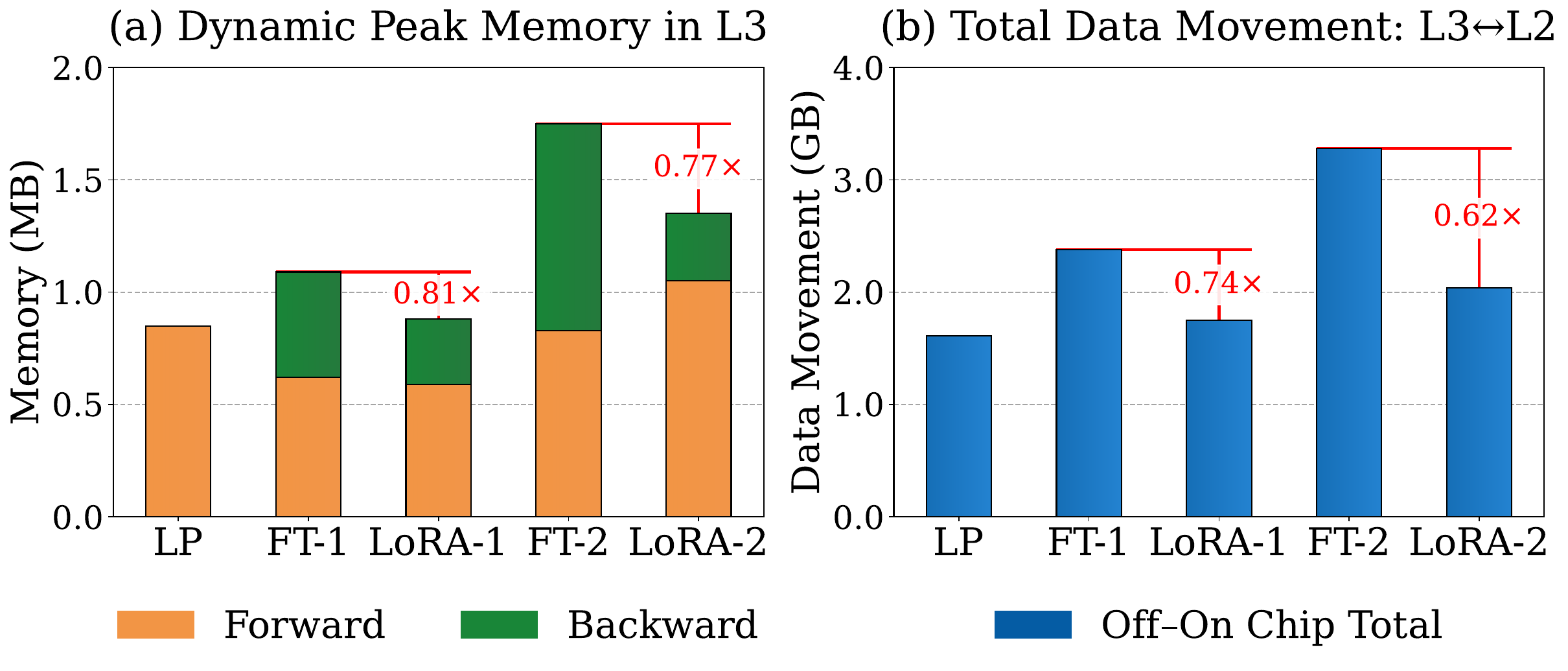}
    \caption{(a) Dynamic peak L3 memory usage during forward and backward passes, 
    including activations and gradients but excluding the fixed 1.12\,MB weights and input. For each configuration, the reported peak memory corresponds to the maximum memory footprint over its training graph lifetime.
    (b) Aggregate off-chip data movement (L3$\leftrightarrow$L2), 
    reported as the total transferred bytes across five strategies.}
    \label{fig:memory}
    \vspace{-0.5cm}
\end{figure}

\Cref{fig:memory} profiles the dynamic peak memory usage and off-chip data movements across the five 
fine-tuning strategies. \Cref{fig:memory} (a) shows the peak dynamic 
allocation in L3, accounting for activations and gradients but excluding weights 
and inputs. The reported values reflect the outcome of allocation 
optimizations performed by the \texttt{MiniMalloc} allocator in Deeploy, where 
the reduced gradients introduced by LoRA are stacked with activations. As 
training depth increases, the dynamic demand rises from less than 1\,MB in LP to nearly 
1.8\,MB in FT\nobreakdash-2, while LoRA lowers this footprint by 19--23\% and reduces the overall off–chip data transfer volume to $0.62\times$ that of full fine-tuning, i.e., a $1.6\times$ reduction, as shown in \Cref{fig:memory}(b).
 
\begin{table*}[t]
  \centering
  \caption{Comparison with related ultra–low-power on-device training work.}
  \label{tab:soa}
  \resizebox{\textwidth}{!}{%
  \begin{threeparttable}
  \begin{tabular}{@{} lllllllll @{}}
    \toprule
    \textbf{System} & 
    \textbf{Hardware} &
    \textbf{Model} &
    \textbf{Memory Layout}\tnote{2} &
    \textbf{Params} &
    \textbf{FW+BW FLOP}\tnote{1} &
    \textbf{FLOP/cyc} &
    \textbf{Training Memory}\tnote{2} \\
    \midrule
    PULP-TrainLib \cite{orailoglu_pulp-trainlib_2022} & PULP-SoC & Deep-AE\,/\,DS-CNN & 64\,KB & 270K\,/\,52.5K & 0.8M\,/\,7.6M & 5.6 & 64\,KB \\
    
    POET \cite{patil_poet_2022} & NRF-52840 & ResNet-18 & 256\,KB + 32\,GB\tnote{3} & 11M & 4.5G & N/A & $<$256\,KB + N/A\tnote{4} \\
    MiniLearn \cite{profentzas_minilearn_nodate} & NRF-52840 & CNNs & 256\,KB + 1\,MB & 79K\,/\,47K\,/\,12K & 1.9M\,/\,2.9M\,/\,0.11M\tnote{5} & 0.59\,/\,1.52\,/\,0.09 & 196\,KB + 439\,KB \\
    TTE\cite{lin_-device_nodate} & STM32F7 & MCUNet & 320\,KB + 1\,MB & 0.48M & $<$69M\tnote{6} & $<$0.43 & 173\,KB \\
    \midrule
    \rowcolor{gray!15}
    \textbf{Ours} & \textbf{PULP-SoC} & \textbf{CCT} & \textbf{128\,KB + 32\,MB} & \textbf{0.28M} & \textbf{71M--126M} & \textbf{4.6} & \textbf{128\,KB + 2.5\,MB} \\
    \rowcolor{gray!10}
    \textbf{Ours} & \textbf{PULP-SoC} & \textbf{Deep-AE} & \textbf{128\,KB + 32\,MB} & \textbf{270K} & \textbf{0.8M} & \textbf{13.4} & \textbf{128\,KB} \\
    \bottomrule
  \end{tabular}
  \begin{tablenotes}[flushleft]
    \footnotesize
    \item[1] FW+BW = forward + backward operations (FLOP).
    \item[2] On-chip SRAM $+$ off-chip memory (e.g., external Flash or DRAM).
    \item[3] POET leverages a 32\,GB SD card/Flash to enable activations to be stored off-chip.
    \item[4] POET’s training scheme reduces peak SRAM usage but increases dependence on external Flash. The exact external Flash footprint has not been reported.
    \item[5] For MiniLearn, convolution layers are pruned to 75\% sparsity, which reduces operations but incurs up to 10\% accuracy drop.
     \item[6] For TTE, FLOPs are reported as an upper bound, since which layers are updated with sparsity or the exact sparsity ratios are not reported.

  \end{tablenotes}
  \end{threeparttable}}

\end{table*}

\vspace{-0.4em}
\subsection{State-of-the-Art Comparison}
We compare our framework with on\nobreakdash-device training approaches targeting ultra\nobreakdash–low\nobreakdash-power platforms. \Cref{tab:soa} summarizes the key differences in hardware targets, model size, FLOP/cycle, and memory footprint. In addition to our Transformer results (\cgls{cct}), we also include a small-network baseline (Deep\nobreakdash-AE) to align with prior \cgls{cnn}/MLP\nobreakdash-centric work.

\paragraph{PULP-TrainLib~\cite{orailoglu_pulp-trainlib_2022}}
PULP-TrainLib is evaluated on the same PULP cluster configuration as ours (8 RISC-V cores with 4 \cglspl{fpu}), but is confined to \cgls{l1}-only execution and therefore limited to small autoencoders and CNNs, e.g., Deep-AE with 270K parameters and 0.8M forward+backward FLOPs or DS-CNN with 52.5K parameters and 7.6M FLOPs. It achieves a higher throughput of 5.6 FLOP/cycle, thanks to its highly optimized computational primitives. By contrast, our CCT fine-tuning involves a substantially larger model with 0.28M parameters and 71–126M FLOPs, i.e., about two orders of magnitude higher compute demand, which necessarily introduces off-chip/on-chip transfer overheads. Even under this heavier load, CCT still sustains 4.6 FLOP/cycle with the aid of the \cgls{redmule} GEMM accelerator. For a fair comparison, we benchmarked the same Deep-AE, which reaches 13.4 FLOP/cycle, $2.4\times$ higher than PULP-TrainLib, highlighting the benefit of offloading GEMM-dominated kernels to the accelerator while maintaining scalability to larger models.

\paragraph{POET~\cite{patil_poet_2022}}
POET relies on paging and checkpointing to trade off SRAM and latency, enabling ResNet-18 training under a 256\,KB SRAM + 32\,GB Flash setup. However, this introduces extensive off-chip traffic and recomputation. The paper does not report end-to-end latency or FLOP/cycle, preventing a direct throughput comparison. In contrast, our CCT training fits within 128\,KB on-chip SRAM + 2.5\,MB external memory, reducing off-chip reliance by more than an order of magnitude while still sustaining 4.6 FLOP/cycle throughput.

\paragraph{MiniLearn~\cite{profentzas_minilearn_nodate} \& TTE~\cite{lin_-device_nodate}}
MiniLearn demonstrates pruning-driven training with CNNs, reporting 0.59–1.52 FLOP/cycle depending on network setup, but at the cost of up to 10\% accuracy degradation. TTE combines pruning and quantization-aware sparsity to fit MCUNet within 320\,KB SRAM and 1\,MB Flash, but reaches less than 0.43 FLOP/cycle\tnote{6}. Compared with these CNN-focused methods, our CCT fine-tuning sustains 4.6 FLOP/cycle (3–10$\times$ higher) without sacrificing accuracy (50-shot 64.9\% Full-BP vs.\ $80.5\%$ LoRA-2 for EuroSAT), as shown in Table~\ref{tab:all_strategies}.

\paragraph{Comparison}
Unlike prior MCU training approaches that trade accuracy for sparsity or rely on paging/recomputation, TrainDeeploy enables Transformer training by jointly addressing memory and compute. Our results show $71$–$126$M forward+backward FLOPs handled within 128\,KB on-chip SRAM + 2.5\,MB external memory, while achieving up to 96\% accuracy with LoRA fine-tuning. Despite the additional overhead of L3–L2 transfers, our framework sustains 4.6 FLOP/cycle thanks to the \cgls{redmule} accelerator, at least $3\times$ more compute-efficient than sparsity-based methods and with lower external memory dependence than paging-based approaches. Moreover, TrainDeeploy is orthogonal to sparsity and pruning techniques, and can incorporate them to further reduce training cost.

\vspace{-0.3em}
\section{Conclusion and Future Work}
\label{sec:conclusion}

This work presented TrainDeeploy, a framework that enables Transformer training on ultra\nobreakdash-low\nobreakdash-power heterogeneous edge devices. Testing on a SoA SoC, using \cgls{lora}, we reduce trainable parameters and stored gradients by $15\times$ and cut off-chip transfers by $1.6\times$. Offloading \cgls{gemm}s to RedMulE yields the first hardware-accelerated LoRA training at the edge, achieving $2.3$\nobreakdash-$3.5\times$ speedups over an 8-core RISC-V cluster. TrainDeeploy provides a robust and unified toolchain for on-device learning at the extreme edge.

\section{Acknowledgment}
This work has received funding from the Swiss State Secretariat for Education, Research, and Innovation (SERI) under the SwissChips initiative.
\bibliographystyle{IEEEtran}
\bibliography{references_final}

\end{document}